\documentclass[12pt]{article}
\usepackage{color}
\usepackage{latexsym}
\usepackage{epsfig,amssymb,euscript, mathrsfs}
\usepackage{amsmath}
\usepackage{bbm}
\usepackage{tensor}
\usepackage{hyperref}
\newsavebox{\ns}
\newsavebox{\dbrane}
\newsavebox{\dbshort}

\def\be{\begin{equation}}
\def\ee{\end{equation}}
\def\bea{\begin{eqnarray}}
\def\eea{\end{eqnarray}}

\newcommand\diff{\mathrm{d}}

\newcommand{\dd}{\mathrm{d}}
\newcommand{\me}{\mathrm{e}}
\newcommand{\ii}{\mathrm{i}}

\newcommand{\Imag}{\mathrm{Im}\, }
\newcommand{\Real}{\mathrm{Re}\, }

\newlength{\sswidth}

\newcommand{\Lform}{\Omega}

\numberwithin{equation}{section}       % equation numbers in each section
\newcommand{\ud}{\,\mathrm{d}}

\newcommand{\taul}{\eta}

\newcommand{\Cz}{C^{(0)}}
\newcommand{\I}{\Imag[\sigma]}
\newcommand{\Rs}{\Real[\sigma]}

\newcommand{\lb}{\left(}
\newcommand{\rb}{\right)}

\newcommand{\bet}{\gamma}
\begin{document}

\begin{titlepage}
\begin{flushright}
{\tt KCL-MTH-16-06}\\
\end{flushright} 
\begin{center}

\today

\vskip 2.3 cm 

\vskip 5mm

{\Large \bf Supersymmetric AdS$_5$ solutions of}

\vskip 5mm

{\Large \bf  type IIB supergravity without D3 branes}

\vskip 15mm

{Christopher Couzens}

\vskip 1cm

\textit{Department of Mathematics, King's College London, \\
The Strand, London, WC2R 2LS,  UK\\}

\end{center}

\vskip 1 cm

\begin{abstract}
\noindent  We analyse the most general bosonic supersymmetric solutions of type IIB supergravity whose metrics are warped products of five-dimensional anti-de Sitter space (AdS$_5$) 
with a five-dimensional Riemannian manifold $M_{5}$, where the five-form flux vanishes, while  all remaining fluxes are allowed to be non-vanishing consistent with SO(4,2) symmetry. 
This completes the program of classifying all supersymmetric solutions of ten and eleven-dimensional supergravity with an  AdS$_5$ factor. We investigate the supersymmetry conditions
in some special cases, and demonstrate how these are satisfied by a solution originally found in \cite{Macpherson:2014eza}, utilising the method of non-Abelian T-duality.
 
\end{abstract}

\end{titlepage}

\pagestyle{plain}
\setcounter{page}{1}
\newcounter{bean}
\baselineskip18pt
\tableofcontents

\section{Introduction}
Via the AdS/CFT correspondence String or M-theory on a supersymmetric background containing an AdS$_{5}$ factor in the metric is expected to be dual to a four-dimensional superconformal field theory \cite{Maldacena:1997re}. As such, there has been much interest in classifying supersymmetric AdS$_{5}$ solutions of IIA and IIB supergravity and M-theory.  In \cite{Gauntlett:2005ww} AdS$_{5}$ solutions of IIB with non-vanishing $F_{5}$ Ramond-Ramomd (R-R) flux were classified. Whilst in \cite{Apruzzi:2015zna} supersymmetric AdS$_{5}$ solutions of massive IIA were classified and new analytic solutions found. An analogous classification for M-theory was carried out in \cite{Gauntlett:2004zh} and many new solutions were found.\footnote{A later refinement of this work was carried out in \cite{Lin:2004nb} in which the additional conditions for $\mathcal{N}=2$ supersymmetry were considered. It was later shown in \cite{OColgain:2010ev} that the classification of \cite{Lin:2004nb} was the most general consistent with $\mathcal{N}=2$ supersymmetry and an AdS$_{5}$ factor in M-theory. A later refinement of \cite{Gauntlett:2005ww} was carried out in \cite{Colgain:2011hb} to impose the additional condition of $\mathcal{N}=2$ supersymmetry.} In this work we plug a remaining gap in the classification of the IIB case. An alternative method for classifying supersymmetric supergravity solutions with an AdS factor in the metric, to that used in the above references and in this paper, was carried out in \cite{Gutowski:2014ova, Beck:2014zda, Beck:2015hpa}. 

The case of vanishing self-dual five form, $F_{5}$, was not considered in \cite{Gauntlett:2005ww} and was implicitly assumed to be non-vanishing throughout. Attempts to set $F_{5}=0$  in the final equations of \cite{Gauntlett:2005ww} run into inconsistencies as it involves dividing by zero. This case of vanishing $F_{5}$ corresponds to having no D3 branes in the theory and there is a close analogy between this and the no M2 branes case of AdS$_{4}$ in eleven-dimensional supergravity which was first classified in \cite{Gauntlett:2006ux} and later extended in \cite{Gabella:2012rc}.

Completing this classification was motivated in part by the recent solutions found in \cite{Macpherson:2014eza}. Two new supersymmetric solutions of IIB supergravity were found with $F_{5}=0$ and are the first of their type. To obtain these solutions the authors begin with two well known AdS$_{5}$ Sasaki-Einstein solutions and perform a Non-Abelian T-duality (NATD) on an $SU(2)$ isometry to IIA followed by a T-duality along a remaining $U(1)$ to return to IIB. The supersymmetric solutions that are obtained have seed solutions AdS$_{5}\times T^{(1,1)}$ and AdS$_{5}\times Y^{p,q}$. Unfortunately these new solutions are singular and it was hoped that by completing this classification we would be able to find new non-singular solutions of this form. Finding non-singular AdS$_{5}$ solutions with vanishing $F_{5}$ remains an open problem.

In this paper we consider the most general bosonic supersymmetric solutions of type IIB supergravity with a warped metric of the form AdS$_{5}$$\times M_{5}$, where $M_{5}$ is an internal manifold that admits a Riemannian metric. We set the self-dual five-form field strength, $F_{5}$, to be vanishing but allow all other Neveu-Schwarz Neveu-Schwarz (NS-NS) and R-R fluxes to be non-vanishing and consistent with preserving the $SO(4,2)$ symmetry of AdS$_{5}$. We use the well known method of analysing the G-structure determined by the Killing spinors as was employed, for example, in \cite{Gauntlett:2005ww} (and references therein) from which some of this work is derived. We find that the internal manifold admits an identity structure which allows us to determine the metric in full generality. The geometry includes a hypersurface-orthogonal Killing vector which is a symmetry of the full solution and corresponds to the $U(1)$ R-symmetry in the putative dual superconformal field theory. Furthermore, analogous to the conclusion in \cite{Gauntlett:2005ww}, we find that supersymmetry implies that all the equations of motion and Bianchi identities are satisfied, though this does not follow immediately from their work. 

The plan for the paper is as follows. In section \ref{SectionSUSY} we present the conditions for preserving supersymmetry. In section \ref{Sectionbieq} we present the torsion conditions and show that supersymmetry implies all the equations of motion and Bianchi identities. In section \ref{SectionLocalCoords} we further the analysis by introducing local coordinates and reduce to a minimal set of necessary and sufficient conditions for a supersymmetric solution. In section \ref{secP=0} we consider a simple ansatz and find a singular solution, in section \ref{secPnot=0} we present a less simplified ansatz and reduce the solution to a single ODE to solve. In section \ref{NATDsection} we show that the NATD-T-dual of AdS$_{5}\times T^{(1,1)}$ solution found in \cite{Macpherson:2014eza} satisfies our equations. We conclude in section \ref{Conclusion}. We relegate some definitions and technical details to three appendices. The first contains the definitions of the bilinears and the calculation of the orthonormal frame used in the paper, the second contains algebraic analysis for the existence of non-singular solutions to the ansatz of section \ref{secP=0}, whilst the third contains technical material used in section \ref{NATDsection}.

%%%%%%%%%%%%%%%%%%%%%%%%%%%%%%%%%%%%%%%%%%%%%%%%%%%%%%%%%%%%%%%%%%%%%%%%%%%%%%%%

\section{The conditions for supersymmetry in $d=5$}\label{SectionSUSY}

We shall follow the conventions and notation of \cite{Gauntlett:2005ww} for the type IIB supergravity field content, equations of motion, and supersymmetry variations. 
In addition to the  ten-dimensional metric $g_{MN}$, the bosonic fields comprise the axion-dilaton $\tau=C^{(0)}+\ii \me^{-\Phi}$, a complex three-form flux  
\begin{gather}
G=\ii \me^{\Phi/2}(\tau \ud B -\ud C^{(2)})~,
\end{gather}
where $B$ and  $C^{(2)}$, are the  NS-NS and R-R two-form potentials, respectively, and a self-dual five-form $F_{5}=*_{10}F_{5}$.  Moreover, the axion and dilaton 
enter the equations of motion and supersymmetry variations through  the following one-forms 
\bea
P&=&\frac{\ii}{2}\me^{\Phi}\ud C^{(0)}+\frac{1}{2}\ud \Phi~,\\
Q&=&-\frac{1}{2}\me^{\Phi}\ud C^{(0)}~.
\eea
The covariant derivative $D_M$ with respect to both local Lorentz transformations and local $U(1)$ gauge transformations, is defined as 
\begin{equation}
D_{M}=\nabla_{M}-\ii q Q_{M}~,
\end{equation}
where $q$ is the charge of the field under the local $U(1)$: $P$ has charge 2, $G$ has charge 1 and the Killing spinor $\epsilon$ has charge $1/2$.
We refer the reader to \cite{Gauntlett:2005ww} for the equations of motion, Bianchi identities, and the supersymmetry variations for the gravitino $\psi_M$ and dilatino $\lambda$.

We wish to characterise the most general class of bosonic supersymmetric solutions of type IIB supergravity with $SO(4,2)$ symmetry and \emph{vanishing five-form flux}.  Namely
we require that
\be
F_{5}=0~,
\ee
which means that the solutions we study correspond to configurations \emph{without D3 branes}. 
This is a slight difference to the analysis performed in \cite{Gauntlett:2005ww}, where it was (implicitly) assumed throughout that $F_{5}\not=0$. As pointed out in the introduction it is not possible to simply set $F_{5}=0$ in the final equations presented in \cite{Gauntlett:2005ww}. Nevertheless much of the initial analysis conducted in their paper can be utilised and we shall indicate when this is possible and when it is not.

The $d=10$ metric, in Einstein frame, takes the form of a warped product
\begin{equation}
\ud s_{10}^{2}=\me^{2\Delta}\left(\ud  s^{2}_{\text{AdS}_{5}}+\ud s_{M_{5}}^{2}\right)\label{metric},
\end{equation}
where $\ud s^{2}_{\text{AdS}_{5}}$ is the metric on AdS$_{5}$ with Ricci tensor given by $ R_{\mu\nu}=-4m^{2}(g_{\text{AdS}_{5}})_{\mu\nu}$
and  $\ud s_{M_{5}}^{2}$ is the metric on a  five-dimensional Riemannian internal space $M_{5}$. In order to preserve the $SO(4,2)$ symmetry of the metric we require the fields to take values in; $\Delta\in \Omega^{0}(M_{5},\mathbb{R})$, $P\in \Omega^{1}(M_{5},\mathbb{C}), Q\in \Omega^{1}(M_{5},\mathbb{R})$ and $G\in \Omega^{3}(M_{5},\mathbb{C})$. Notice that with this ansatz the Bianchi identity for $F_{5}$ is trivially satisfied and it is therefore consistent to set $F_{5}=0$ without imposing any further conditions.

We will use the most general ansatz for the Killing spinor consistent with preserving minimal supersymmetry in AdS$_{5}$. This takes the form
\begin{equation}
\epsilon=\me^{\Delta/2}(\psi\otimes \xi_{1}\otimes \theta+\psi^{c}\otimes \xi_{2}^{c}\otimes \theta)~,
\end{equation}
where we have rescaled the spinor by the factor $\me^{\Delta/2}$ for later convenience. Here $\psi$ is a Killing spinor on AdS$_5$ and $\xi_i$ are two independent 
$Spin(5)$ spinors on $M_5$. Further discussion about the spinor ansatz and conventions can be found in appendix A of \cite{Gauntlett:2005ww}. Requiring supersymmetry to be preserved
 yields the following  conditions 
\bea
D_{m}\xi_{1}+\frac{1}{8}\me^{-2\Delta}\gamma^{m_{1}m_{2}}G_{mm_{1}m_{2}}\xi_{2}-\frac{\ii}{2}m\gamma_{m}\xi_{1}&=&0~, \label{fd1}\\
\bar{D}_{m}\xi_{2}+\frac{1}{8}\me^{-2\Delta}\gamma^{m_{1}m_{2}}G_{mm_{1}m_{2}}^{*}\xi_{1}-\frac{\ii}{2}m\gamma_{m}\xi_{2}&=&0~,\label{fd2}\\
\gamma^{m}\partial_{m}\Delta \xi_{1}+\ii m\xi_{1}-\frac{1}{48}\me^{-2\Delta}G_{m_{1}..m_{3}}\gamma^{m_{1}..m_{3}}\xi_{2}&=&0~,\label{fa1}\\
\gamma^{m}\partial_{m}\Delta \xi_{2}+\ii m\xi_{2}-\frac{1}{48}\me^{-2\Delta}G^{*}_{m_{1}..m_{3}}\gamma^{m_{1}..m_{3}}\xi_{1}&=&0~,\label{fa2}\\
P_{m}\gamma^{m}\xi_{2}+\frac{1}{24}\me^{-2\Delta}\gamma^{m_{1}..m_{3}}G_{m_{1}..m_{3}}\xi_{1}&=&0~,\label{fa3}\\
P_{m}^{*}\gamma^{m}\xi_{1}+\frac{1}{24}\me^{-2\Delta}\gamma^{m_{1}..m_{3}}G^{*}_{m_{1}..m_{3}}\xi_{2}&=&0\label{fa4}~.
\eea
These can be obtained straightforwardly from the equations (3.3) - (3.8) in  \cite{Gauntlett:2005ww}, by setting $f=0$\footnote{$f$ is the constant defined in \cite{Gauntlett:2005ww} as $F_{5}=f (\text{Vol}_{\text{AdS}_{5}}+\text{Vol}_{5})$.}.

\subsection*{Special cases}

The possible stabilizer groups of the Spin(5) spinors $\xi_{i}$ are the identity group or $SU(2)$. Consequently  $M_{5}$ may admit either an identity structure or an $SU(2)$ structure.

Let us first consider the case of an $SU(2)$ structure. This corresponds to setting one of the spinors to zero, without loss of generality, let us assume  $\xi_{2}=0$. Then equation (\ref{fa1}) reads
\begin{equation}
\gamma^{m}\partial_{m}\Delta\xi_{1}=-\ii m \xi_{1}~.\label{fa1xi2=0}
\end{equation}
Following the use of Clifford algebra identities   one can show easily that $\partial_{n}\Delta=0$, and inserting this back into (\ref{fa1xi2=0}) we reach the contradiction $m \xi_{1}=0$. Whilst the $F_{5}\not=0$ case allowed for an $SU(2)$ structure on $M_{5}$, comprising  the well known Sasaki-Einstein solutions, we conclude that there are no supersymmetric AdS$_{5}\times M_{5}$ solutions with $F_{5}=0$ in type IIB supergravity with $M_{5}$ admitting an $SU(2)$ structure\footnote{In  \cite{Apruzzi:2015zna}
 it has also been shown that in type IIA supergravity there are no solutions of the form AdS$_{5}\times M_{5}$ with $M_{5}$ having an $SU(2)$ structure either.}.

 Another interesting case to consider is $G=0$. Such putative solutions would arise purely from D7 branes, and would be motivated by F-theory constructions. Setting $G=0$ in equation (\ref{fa1}) and (\ref{fa2}) once again gives (\ref{fa1xi2=0}) and an analogous equation for $\xi_{2}$ which implies $\xi_{1}=0=\xi_{2}$ and hence no supersymmetry is preserved. 
 We therefore conclude that supersymmetric AdS$_5$ solutions of type IIB supergravity with vanishing five-form \emph{and} three-form fluxes  do not exist.
 
 In the remainder of the paper we will assume that $G$ is non-vanishing, and that both spinors $\xi_i$ are not identically zero, thus giving a (local) identity structure on $M_5$.

\section{Bilinear equations}\label{Sectionbieq}

The identity structure is characterised by a set of one-forms, constructed as spinor bilinears, that can be used to define a canonical orthonormal frame on $M_5$. In the analysis of the algebraic and differential conditions equivalent to the supersymmetry equations it is useful to consider also a number of scalar and two-form bilinears. We define these following the notation 
 in \cite{Gauntlett:2005ww} and we list them in appendix \ref{Oframe}. From the algebraic condition (3.25) in \cite{Gauntlett:2005ww} we see that $F_{5}=0$ implies that 
 $\sin\zeta=0$\footnote{Following the argument in appendix C of \cite{Gauntlett:2005ww}, and imposing $\sin \zeta=0$, we find that it is not possible to have the spinors $\xi_{i}$ non-vanishing and 
 linearly dependent. We therefore restrict to the case of them being independent and admitting an identity structure.}; we can therefore import 
 the bilinear equations from \cite{Gauntlett:2005ww} where we set $\sin\zeta\equiv 0$ and $f\equiv0$. The resulting  differential conditions are\footnote{Here and in the rest of the paper $*$ denotes the Hodge star operator with respect to the five-dimensional metric $\diff s^2_{M_5}$.}
\bea
\me^{-4\Delta}\ud (\me^{4\Delta}S)&=&3\ii m K\label{Seq}~,\\
\me^{-6\Delta}D(\me^{6\Delta}K_{3})&=&P\wedge K_{3}^{*}-4\ii m W -\me^{-2\Delta}*G\label{K3}~ ,\\
\me^{-4\Delta}\ud (\me^{4\Delta}K_{4})&=&-2mV\label{K4}~,\\
 \me^{-8\Delta}\ud(\me^{8\Delta}K_{5})&=&-6mU\label{K5}~,
 \eea
while  the algebraic conditions are
\bea
Z=0&=&\sin \zeta,\quad A=1~,\\
 2i_{K_{3}}\ud\Delta&=&i_{K^{*}_{3}}P~,\\
 i_{K_{5}}\ud \Delta=0&=& i_{K_{5}}P\label{contK5}~,\\
 (1-|S|^{2})\me^{-2\Delta}*G&=&2P\wedge K_{3}^{*}-(4\ud \Delta +4\ii m K_{4})\wedge K_{3}\nonumber\\&&+2*(P\wedge K_{3}^{*}\wedge K_{5}-2\ud \Delta\wedge K_{3}\wedge K_{5})~.
 \label{Geq}
 \eea

Note that  in \cite{Gauntlett:2005ww} the differential condition on $K_4$ was implied by the remaining ones, because this one-form could be expressed as a linear combination of the other bilinears, as can be seen from (\ref{K5alg}), however this is no longer the case. Indeed, more generally, the orthonormal frame that we will use  here, differs from the analogous one introduced in \cite{Gauntlett:2005ww}.
Using this orthonormal frame, presented in  appendix \ref{Oframe}, we find that the metric takes the form

\begin{gather}
 \ud s^{2}_{M_5}=\frac{K_{5}^{2}}{|S|^{2}}+\frac{K_{4}^{2}}{1-|S|^{2}}+\frac{K_{3}\otimes K_{3}^{*}}{1-|S|^{2}}+\frac{|S|^{2}}{1-|S|^{2}}(\mathfrak{Im}[S^{-1}K])^{2}~.\label{Classmetric}
 \end{gather}
This should be contrasted with the metric written in equation  (3.53) of  \cite{Gauntlett:2005ww}.

It is immediate from the analysis of \cite{Gauntlett:2005ww} that $K_{5}$ defines a Killing vector. Moreover, here we will find that  additionally $K_{5}$ is in fact a hypersurface-orthogonal Killing vector.
This is most easily seen after we introduce local coordinates in the following section.

Analogously to \cite{Gauntlett:2005ww}, one can show $K_{5}$ is in fact a symmetry of the full solution, namely
\bea
\mathcal{L}_{K_{5}}\Delta&=&\mathcal{L}_{K_{5}}\Phi=\mathcal{L}_{K_{5}}C^{(0)}=0~,\nonumber\\
\mathcal{L}_{K_{5}} G&=&0~.
\eea
 In a putative dual $d=4$ superconformal field theory this corresponds to having $U(1)$ R-symmetry and hence $\mathcal{N}=1$ supersymmetry.

Let us  now show that supersymmetry implies that all the equations of motion and Bianchi identities are satisfied.
Most of the arguments presented in \cite{Gauntlett:2005ww} to show that all the equations of motion and the $P$ Bianchi identity are implied by supersymmetry can be used  in our case, however, as alluded to in the introduction the argument showing that the Bianchi identity for $G$ is satisfied is not valid if $F_{5}=0$.  Below we present an argument that applies to both cases. Using the supersymmetry equations, we find
\bea
D(\me^{6\Delta}X)&=&\me^{6 \Delta}(3\ii m *X-\me^{-2 \Delta}S G+P\wedge Y) \label{DX}~,\\
\me^{-6\Delta}\bar{D}(\me^{6\Delta}Y)&=&3 \ii m *Y+\me^{-2\Delta}S G^{*}+P^{*}\wedge X~,\label{DY}\\
\me^{-6\Delta}D(\me^{6\Delta}*X)&=&-\me^{-2\Delta}G\wedge K+P\wedge *Y \label{D*X}~.
\eea
These equations are true even including a non-zero $F_{5}$, as this drops out of the expressions.
To recover the Bianchi identity for $G$ one should take $D$ of (\ref{DX}) and use (\ref{Seq}), (\ref{DY}) and (\ref{D*X}). 
As in \cite{Gauntlett:2005ww}, we  conclude:
\begin{quote}
\emph{For the class of solutions with metric of the form (\ref{metric}), vanishing five-form flux and fluxes respecting $SO(4,2)$ symmetry, all the equations of motion and Bianchi identities are implied by supersymmetry.}
\end{quote}

\section{Introducing local coordinates}\label{SectionLocalCoords}

In this section we shall introduce  local coordinates in which the set of BPS  equations become more explicit. 
We begin by  reducing on the Killing direction defined by $K_{5}$, resulting in a 4-1 splitting of the metric. 
The transverse four-dimensional metric to the Killing direction admits an integrable almost product structure giving a further $3$-$1$ splitting. 
The resulting BPS equations take a similar form to those presented in  \cite{Gauntlett:2005ww} in  the $F_{5}\not=0$ case, but they are different. 
We shall conclude this section by introducing explicit coordinates on the remaining three-dimensional part of the metric, and 
obtaining expressions for the NS-NS and R-R two-form potentials.

 We begin by choosing a local coordinate  adapted to the Killing direction defined by $K_{5}$. As a vector we have
 \begin{gather}
 K_{5}^{\#}=3m\frac{\partial}{\partial\psi}~,
\end{gather}
and as a one-form
\begin{equation}
 K_{5}=\frac{|S|^{2}}{3m}(\dd \psi+\rho)~,
 \end{equation}
where $\rho$ is a one-form with no $\ud\psi$ term. The factor of $3m$ is chosen for later convenience. The Lie derivative of $S$ with respect to $K_{5}^{\#}$ is
\begin{gather}
\mathcal{L}_{K_{5}^{\#}}S=-3 \ii m S~,
\end{gather}
from which we find 
\begin{equation}
S=-|S|\me^{-i \psi}~.
\end{equation}
It is convenient to make the redefinitions
\begin{equation}
 \mu=\me^{-4\Delta}~,\quad \taul =\me^{4\Delta}|S|~.
\end{equation}
Then from (\ref{Seq}) we have
\begin{equation}
 K=\frac{\mu~  \me^{-\ii \psi}}{3m}(\taul\ud \psi+\ii  \ud \taul)~,
 \label{explicitK}
 \end{equation}
and using the expression for $K$ in appendix \ref{Oframe} we deduce that 
 \begin{equation}
K_{5}=\frac{\taul^{2}\mu^{2}}{3m}\ud \psi~,\label{K5final}
\end{equation}
and is therefore a hypersurface-orthogonal Killing vector. Notice that the Killing vector is not fibered, $\rho=0$, and this differs from \cite{Gauntlett:2005ww}. Making the additional redefinitions
\begin{equation}
K_{3}=\frac{\mu^{3/2}}{3m}\sigma~,\quad K_{4}=\frac{\mu}{3m} \beta~,\label{K3final}
\end{equation}
the metric becomes
\begin{equation}
 9m^{2}\ud s^{2}=\frac{1}{1-\taul^{2}\mu^{2}}\left(\mu^{3}\sigma\otimes \sigma^{*}+\mu^{2}\beta^{2}+\mu^{2}\ud \taul^{2}\right) + \taul^{2} \mu^{2}\ud \psi^{2}~.\label{localmetric}
 \end{equation}

Here $\beta$ is a real one-form and  $\sigma$  is a complex one-form, and both have no leg along the Killing direction. 
We should  now re-express the differential and algebraic conditions in terms of these redefined quantities. We find that (\ref{K5}) is 
automatically satisfied, whilst equation (\ref{K4}) becomes
 \begin{gather}
 \ud \beta=\frac{\mu^{2}}{3(1-\taul^{2}\mu^{2})}\left[ \ii \sigma^{*}\wedge \sigma-2\taul \ud \taul \wedge \beta\right]~.
 \label{beta}
\end{gather}
 Equation (\ref{K3}) becomes
 \begin{gather}
D\sigma=\frac{1}{\taul^{2}\mu^{2}-1}\left[(1+\taul^{2}\mu^{2})P\wedge \sigma^{*}+\frac{4\mu^{2}\taul}{3}\ud \taul \wedge \sigma +\ud \ln \mu\wedge \sigma\right. \nonumber\\[2mm]%%%
\left.+\frac{\taul^{2}\mu^{2}}{3m}*\left(2P\wedge \sigma^{*}\wedge \ud \psi+\ud\ln \mu\wedge \sigma\wedge \ud \psi\right)\right]~,\label{sigma}
\end{gather}
 where we have used the expression for $*G$ given in (\ref{K3}).
The remaining algebraic equations read
\begin{gather}
2 i_{\sigma^{*}}P=-i_{\sigma} \ud \ln\mu~, \label{alg} \\
\mathcal{L}_{\frac{\partial}{\partial\psi}}\mu=\mathcal{L}_{\frac{\partial}{\partial\psi}}\Phi=\mathcal{L}_{\frac{\partial}{\partial \psi}}C^{0}=0~.
\label{psicomp}
\end{gather}
These constitute the set of necessary and sufficient conditions that one needs to satisfy for supersymmetry.

To make these equations completely explicit, we can introduce the four remaining coordinates. It is a standard calculation (for example starting with (\ref{explicitK})) to check that the four-dimensional metric transverse to the Killing direction has an integrable almost product structure. This allows one to introduce  ``splitting coordinates'', and gives a 3-1 splitting of the metric. In these coordinates the metric still takes the form presented in (\ref{localmetric}) however now the one-forms $\beta$ and $\sigma$ have no $\ud \taul$ term, though they are still in general functions of $\taul$. 
 We may then split the five-dimensional exterior derivative as
\begin{equation}
\ud =\ud_{3}+\ud \taul \frac{\partial}{\partial\taul}+\ud \psi \frac{\partial}{\partial \psi}~,
\end{equation}
where $\ud_{3}$ is the exterior derivative on the three-dimensional metric defined by the integrable almost product structure.
Equation (\ref{beta}) now reads
\bea
\ud_{3}\beta&=&\frac{\ii \mu^{2}}{3(1-\taul^{2}\mu^{2})} \sigma^{*}\wedge \sigma~,\label{beta3eq}\\
\partial_{\taul}\beta&=&-\frac{2 \taul\mu^{2}}{3(1-\taul^{2}\mu^{2})} \beta~, \label{betataueq}
\eea
whilst (\ref{sigma}) reads\footnote{Here $*_{3}$ is the hodge star on the three-dimensional metric defined by the integrable almost product structure.}
\bea
\ud_{3} \sigma-\ii Q_{3} \wedge \sigma&=&\left.\frac{1}{\taul^{2}\mu^{2}-1}\right[ (1+\taul^{2}\mu^{2})P_{3}\wedge \sigma^{*}+\ud_{3}\ln\mu \wedge \sigma \nonumber\\%%%%%
 &&\left.-3 m\taul \sqrt{1-\taul^2 \mu^2}*_{3}\left(2P_{\taul}\sigma^{*}+\partial_{\taul}\ln\mu \sigma\right)\right]~,\label{d3sigmaeq}\\%%%%%%
\partial_{\taul}\sigma-\ii Q_{\taul}\sigma&=&
\frac{1}{\taul^{2}\mu^{2}-1}\left[ (1+\taul^{2}\mu^{2})P_{\taul}\sigma^{*}+\frac{4 \mu^{2}\taul}{3}\sigma+\partial_{\taul}\ln\mu~ \sigma\right.\nonumber\\%%%%%%%
&&\left.-\frac{ \mu^2 \taul}{ 3m \sqrt{1-\taul^2 \mu^2}}*_{3}\left(2P_{3}\wedge \sigma^{*}+\ud_{3}\ln \mu \wedge \sigma\right)\right]~,
\label{sigmataueq}
\eea
where we have used (\ref{psicomp}). 

Thus for the most general, minimally supersymmetric AdS$_{5}$ solutions with vanishing five-form flux we need to solve the four differential equations
 (\ref{beta3eq}) - (\ref{sigmataueq}) subject to the algebraic equation 
 (\ref{alg}). 
 We note that the integrability equation for (\ref{beta3eq}) and (\ref{betataueq}) is automatically satisfied upon using  (\ref{alg}), (\ref{d3sigmaeq}) and (\ref{sigmataueq}). 

 We may now introduce the three remaining coordinates along $\beta$ and $\sigma$, which we will denote as  $x$ and $y_{i}$, with $i=1,2$.
 In particular,  we write the three independent real one-forms as
\bea
\beta&=&\gamma_{x}\ud x+\gamma_{y_{1}}\ud y_{1}+\gamma_{y_{2}}\ud y_{2}~,\nonumber\\
\Real[\sigma]&=&\rho_{x}\ud x+\rho_{y_{1}}\ud y_{1}+\rho_{y_{2}}\ud y_{2}~,\label{3coords}\\
\Imag[\sigma]&=&\kappa_{x}\ud x+\kappa_{y_{1}}\ud y_{1}+\kappa_{y_{2}}\ud y_{2}~.\nonumber
\eea
Notice that generically we cannot simplify further these expressions, and the equations (\ref{beta3eq}) - (\ref{sigmataueq}) take the form of a very complicated set of coupled PDE's.
 An explicit example of a rather generic solution will be presented later in section \ref{NATDsection}.

To obtain the explicit form of the NS-NS two-form $B$ and the R-R two-form $C^{(2)}$ we can combine equations (\ref{DX}) and (\ref{DY}), 
to obtain
\bea
D (\me^{6\Delta}(Y^{*}-X))= -3 \ii m \me^{6\Delta}*(Y^{*}+X)+\me^{4\Delta}(S+S^{*})G+\me^{6\Delta}P\wedge(X^{*}-Y)~.
\label{tofindB}
\eea
It is then simple, but tedious, to extract the two two-forms $B$ and $C^{(2)}$ from the real and imaginary parts of this equation, by using  (\ref{Seq}) - (\ref{K5}) and the results of appendix \ref{Oframe}.
We find
\bea
B-\omega_{B}&=&\frac{\me^{\Phi/2}\mu}{9m^{2}}\Real[\sigma]\wedge \ud \psi ~,        \label{Beq}\\
C^{(2)}-\omega_{C}&=&C^{(0)}\frac{\me^{\Phi/2}\mu}{9m^{2}}\Real[\sigma]\wedge \ud \psi+\frac{\me^{-\Phi/2}\mu}{9m^{2}}\Imag[\sigma]\wedge \ud \psi  ~,               \label{C2}
\eea
where $\omega_{B}$ and  $\omega_{C}$ are undetermined closed two-forms. Analogous expressions relevant for the $F_{5}\neq 0$ case were given in  \cite{Gabella:2009wu}. 

%%%%%%%%%%%%%%%%%%%%%%%%%%%%%%%%%%%%%%%%%%%%%%%%%%%%%%%%%%%%%%%%%%%%%%%%

\section{Complex $M_4$ and $P=0$}\label{secP=0}

Motivated by finding explicit solutions we set $P=0$ in this section\footnote{This condition imposes that the distinguished transverse four-dimensional foliation defined by the Killing vector $\partial_{\psi}$, which we call $M_{4}$, has an integrable almost complex structure. Consider a holomorphic two-form constructed from the orthonormal frame of appendix \ref{Oframe} as
\bea
\Lform&\equiv &(e^{2}+\ii e^{5})\wedge (e^{4}-\ii e^{3})~\nonumber\\[2mm]
&=& \frac{1}{2(\taul \mu-1)}(\me^{ \ii \psi} X+\me^{- \ii \psi} Y^{*}+2 W)~.
\label{defLform}
\eea
This then defines an almost complex structure on $M_4$. In the second line we have expressed $\Omega$ in terms of the two-form bilinears. Imposing that this is integrable implies 
\begin{equation}
P=g(e^{4}+\ii e^{3})+f(e^{2}+\ii e^{5})+ h(e^{4}-\ii e^{3})~,
\end{equation}
where $f,g,h$ are arbitrary complex functions (subject to satisfying the $P$ equation of motion and Bianchi identity). Setting $P=0$ solves this constraint therefore $M_{4}$ is complex in this case. It would have been more interesting to impose this more general form of $P$, however it is still a fairly complicated system of equations to solve and we were unable to do so.}. Notice that setting $P=0$ implies that $\mu$ is a function of $\taul$ only\footnote{To see this use (\ref{alg}) to note that $\ud \ln\mu= f_{K_{4}} K_{4}+f_{\taul}\ud \taul$ for some real functions $f_{K_{4}}$ and $f_{\taul}$. Requiring that this is closed then implies that $f_{K_{4}}=0$.}. Setting $P=0$ and $\mu=\mu(\taul)$ reduces the necessary and sufficient differential equations to 
\bea
\ud_{3} \beta&=&\frac{2 \mu^{2}}{3(1-\taul^{2}\mu^{2})} \I \wedge \Rs\label{2cdb}~,\\
\ud_{3}\Rs&=&\frac{\mu \taul}{\taul^{2}\mu^{2}-1}\partial_{\taul}\ln \mu~ \beta \wedge \I\label{2cdr}~,\\
\ud_{3}\I&=&-\frac{\mu \taul}{\taul^{2}\mu^{2}-1}\partial_{\taul}\ln \mu~ \beta \wedge \Rs\label{2cdi}~,
\eea
and
\bea
\partial_{\taul}\beta&=&-\frac{2 \taul \mu^{2}}{3(1-\taul^{2}\mu^{2})}\beta  \label{2ctb}~,\\
\partial_{\taul}\Rs&=&\frac{1}{\taul^{2}\mu^{2}-1}\left(\frac{4 \mu^{2}\taul}{3}+\partial_{\taul}\ln \mu\right)\Rs ~,\label{2ctr}\\
\partial_{\taul}\I&=&\frac{1}{\taul^{2}\mu^{2}-1}\left(\frac{4 \mu^{2}\taul}{3}+\partial_{\taul}\ln\mu\right)\I    \label{2cti}~. 
\eea
We see immediately that we may solve (\ref{2ctb}) - (\ref{2cti}) as
\bea
\beta&=&\exp\left[\int\frac{2 \mu^{2} \taul}{3(\taul^{2}\mu^{2}-1)}\ud \taul \right]\hat{\beta}\label{taubeta}~,\\
\Rs&=&\exp\left[\int\frac{1}{\taul^{2}\mu^{2}-1}\left(\frac{4 \mu^{2}\taul}{3}+\partial_{\taul}\ln \mu\right)\ud \taul\right] \hat{R}~,\label{tauhatR}\\
\I&=&\exp\left[\int\frac{1}{\taul^{2}\mu^{2}-1}\left(\frac{4 \mu^{2}\taul}{3}+\partial_{\taul}\ln \mu\right)\ud \taul\right] \hat{I}~,\label{tauhatI}
\eea
where the hatted objects are $\taul$ independent one-forms. We note that the above integrations may include arbitrary integration constants which we absorb into the $\taul$ independent one-forms. Upon substituting these expressions into (\ref{2cdb}) - (\ref{2cdi}) one sees that the $\taul$ dependence in (\ref{2cdb}) cancels automatically as it should. However the $\taul$ dependence in (\ref{2cdr}) and (\ref{2cdi}) does not, we should have been suspicious if it cancelled as it would imply that $\mu$ could be any function of $\taul$, requiring that this expression is $\taul$ independent gives us the defining differential equation for $\mu$
\begin{equation}
\partial_{\tau}\left(\frac{\mu \taul}{\taul^{2}\mu^{2}-1} \partial_{\taul}\ln\mu \exp\left[\int\frac{2 \mu^{2} \taul}{3(\taul^{2}\mu^{2}-1)}\ud \taul \right]\right)=0~.
\end{equation}
We find a solution to the system of differential equations if we satisfy the second order non-linear differential equation
\begin{equation}
(3+\taul^{2}\mu^{2})\dot{\mu}+6 \taul^{3} \mu \dot{\mu}^{2}+3 \taul (1-\taul^{2}\mu^{2})\ddot{\mu}=0~,   \label{secordermu}
\end{equation}
and the three differential equations
\bea
\ud_{3}\hat{\beta}&=&\frac{2}{3}\hat{I}\wedge \hat{R}~,\label{hatbeta}\\
\ud \hat{R}&=&c \hat{\beta}\wedge \hat{I}~,\label{hatR}\\
\ud \hat{I}&=&-c \hat{\beta} \wedge \hat{R}~.\label{hatI}
\eea
Where $c$ is a constant satisfying
\begin{equation}
c=\frac{\mu \taul}{\taul^{2}\mu^{2}-1} \partial_{\taul}\ln\mu \exp\left[\int\frac{2 \mu^{2} \taul}{3(\taul^{2}\mu^{2}-1)}\ud \taul \right]~.
\end{equation}
Notice that $c$ is non-zero if $\mu$ is non-constant and we shall distinguish between these two cases. For the $c=0$ case we can write the solution in closed form and we will discuss it in the remainder of this section. However we are unable to write the $c \not=0$ case in closed form and instead present algebraic analysis for the existence of non-singular solutions in appendix \ref{Appanalysis}.

%%%%%%%%%%%

\subsection*{A singular solution}
We look at the $c=0$ solution  of (\ref{secordermu}) which is equivalent to constant $\mu$. For simplicity we set $\mu=1$. We are now able to integrate (\ref{taubeta})- (\ref{tauhatI});  we find
\bea
\beta&=&(1-\taul^{2})^{1/3}\hat{\beta}~,\qquad \partial_{\taul}\hat{\beta}=0~,\\
\Rs&=&(1-\taul^{2})^{2/3}\hat{R}~,\qquad \partial_{\taul}\hat{R}=0~,\\
\I&=&(1-\taul^{2})^{2/3}\hat{I}~,~\qquad \partial_{\taul}\hat{I}=0~.
\eea
We then need to solve
\bea
\ud_{3}\hat{R}&=&0~=~\ud_{3}\hat{I}~,\\
\ud_{3}\hat{\beta}&=&\frac{2}{3}\hat{I}\wedge \hat{R}  \label{betaconstsol}~.
\eea
As $\hat{R}$ and $\hat{I}$ are closed we may define coordinates $y_{1}$ and $y_{2}$ such that
\begin{equation}
\hat{R}=\ud y_{2}~,\quad \hat{I}=\ud y_{1}~.
\end{equation}
A solution to (\ref{betaconstsol})  is 
\begin{gather}
\hat{\beta}=\frac{2}{3}(\dd x + y_{1} \dd y_{2})~.
\end{gather}
The metric is 
\bea
9 m^2 \ud s^{2}&=&\taul^{2}\ud \psi^{2}+(1-\taul^{2})^{1/3} (\ud y_{1}^{2}+ \ud y_{2}^{2})\nonumber\\
&&+\frac{4}{9(1-\taul^{2})^{1/3}}(y_{1} \ud y_{2}+\ud x)^2+\frac{1}{1-\taul^{2}}\ud \taul^{2}~,
\eea
and we have
\bea
B&=&\frac{(1-\taul^{2})^{2/3}}{9m^2}\ud y_{2}\wedge \ud \psi ~,\\
C^{(2)}&=&\frac{(1-\taul^{2})^{2/3}}{9m^{2}}\ud y_{1}\wedge \ud \psi ~.
\eea
Note that the range of $\taul$ should be either $\taul \in [0,1]$ or $\taul\in [-1,0]$. We find that the Ricci Scalar is given by $R=28 m^{2}$, whilst $R_{\mu\nu}R^{\mu\nu}=336 m^4$ however we find that $R_{\mu_{1}..\mu_{4}}R^{\mu_{1}..\mu_{4}}$ exhibits a singularity as $\taul\rightarrow \pm 1$ and therefore the solution is singular.\\
We note that for $F_{5}\not=0$ an analogous solution of the equations of \cite{Gauntlett:2005ww} exists, which was missed previously, by setting $\phi,\Cz$ and the warp factor to be constants. This solution is once again singular and the singularity appears first in the Ricci scalar, it has non-zero $G$ and hence is also not Sasaki-Einstein. These solutions are unusual in the sense that the only other known solutions with constant warp factor are the Sasaki-Einstein solutions.

\section{An ansatz with $P\neq 0$}\label{secPnot=0}

The structure of the BPS equations suggests an ansatz in which the $\taul$ coordinate plays a distinguished role, therefore we make an ansatz where everything depends non-trivially on this coordinate only. This ansatz is also motivated by the existence of analogous solutions of other BPS systems.  More concretely, we can attempt an ansatz precisely analogous to the one used in section 5 of  \cite{Gauntlett:2005ww} which led to an ODE for one function with a solution corresponding to the Pilch-Warner solution \cite{Pilch:2000ej}, however the analysis of section \ref{secP=0} suggests that we should relax the assumption $P=0$.

In fact we take a more general ansatz than that considered in \cite{Gauntlett:2005ww} by adding an $SO(2)$ rotation of $\sigma$ by a $\taul$ dependent phase $\theta$. Namely, we consider 
\bea
\beta&=&A(\taul) \tau_{3}~,\\
\sigma&=&\frac{1}{\sqrt{\mu (\taul)}}\me^{\ii\theta (\taul)} (C(\tau)\tau_{2}-\ii B(\tau) \tau_{1})~.
\eea
Where $\tau_a$ are the  $SU(2)$ left-invariant one-forms satisfying $\ud \tau_{1}=\tau_{2}\wedge \tau_{3}$ and cyclic permutations. Here the $\taul$ dependent functions $A,B,C$ are all real valued functions of $\taul$ only.

The part of the BPS system decoupled from the dilaton with respect to the case $\theta=0$, is
\bea
\partial_{\taul}\log B&=& -\frac{1+\taul^{2}\mu^{2}}  {2  \mu \taul} \frac{B}{AC} - \frac{ 4\mu^{2} \taul}{3(1-\taul^{2}\mu^{2})} \label{logBeq1again}\\
\partial_{\taul}\log C&=& -\frac{1+\taul^{2}\mu^{2}}  {2  \mu \taul} \frac{C}{AB} - \frac{ 4\mu^{2} \taul}{3(1-\taul^{2}\mu^{2})}\label{logCeq2again}\\
\partial_\taul \log A&=&-\frac{2 \taul \mu^{2} }{3(1-\taul^{2}\mu^{2})}\label{Aeq3again}\\
A&=&-\frac{2\mu B C}{3(1-\taul^{2}\mu^{2})} \label{Aeq4again}\\
\partial_{\taul} \mu&=&\frac{1-\taul^{2}\mu^{2}}{2\taul A}\left(\frac{C}{B}+\frac{B}{C}\right)\label{mueq5again}~.
\eea
These are four differential equations plus one algebraic, for the four functions $A,B,C,\mu$. However (\ref{Aeq3again}) is redundant and implied by the others, so it can be eliminated to give four equations for four functions, which is encouraging for the existence of solutions. This is a complicated system of ODEs. A possible strategy to solve it is to obtain an ODE of higher degree for one single function; as $\mu$ appears ``most often'' in the system the simplest equation to derive is one for $\mu$. To this end we take two further derivatives of (\ref{mueq5again})
\bea
\dot{\mu}&=&\frac{1-\taul^{2}\mu^{2}}{2 A B C \taul}(C^{2}+B^{2})=-\frac{3(1-\taul^{2}\mu^{2})^{2}}{4 \mu \taul B^{2}C^{2}}(C^{2}+B^{2})~,\\%%
\ddot{\mu}&=&-\frac{3+\taul^{2}\mu^{2}}{3\taul(1-\taul^{2}\mu^{2})}\dot{\mu}-\frac{1+3 \taul^{2}\mu^{2}}{\mu(1-\taul^{2}\mu^{2})}(\dot{\mu})^{2}-\frac{3(1-\taul^{2}\mu^{2})(1+\taul^{2}\mu^{2})}{\mu^{2}\taul(B^{2}+C^{2})}\dot{\mu}~,
\eea
and using the other equations we eventually arrive at the following third order equation\footnote{Note that $\sqrt{3}/\taul$ is a solution to this equation however it gives a metric with incorrect signature and so is discarded.  }
\begin{gather} 
\dddot{\mu}=-\frac{1}{9 \mu \taul^{2} (1-\taul^{2} \mu^{2})^{2}(1+\taul^{2}\mu^{2})}\left[3 \taul(9+47 \taul^{2} \mu^{2}+31 \taul^{4} \mu^{4}+9 \taul^{6} \mu^{6})\dot{\mu}^{2}\right.\nonumber\\
+36 \taul^{4} \mu(2+3 \taul^{2} \mu^{2}+3\taul^{4}\mu^{4})\dot{\mu}^{3}+9\taul \mu(3-3 \taul^{2}\mu^{2}+\taul^{4}\mu^{4}-\taul^{6}\mu^{6}) \ddot{\mu}\nonumber\\%%
+\mu(9+15 \taul^{2}\mu^{2}+35 \taul^{4}\mu^{4}+5 \taul^{6}\mu^{6})\dot{\mu}+9\taul^{2}(3+5\taul^{2}\mu^{2}+\taul^{4}\mu^{4}-9\taul^{6}\mu^{6})\ddot{\mu}\dot{\mu}\left.\right]\label{thirdmu}~.
\end{gather}
One can check that (\ref{secordermu}) actually implies this equation as it should, being the general equation for $P=0$. (\ref{thirdmu}) is clearly a necessary condition for a solution however it is not sufficient, notice that constant $\mu$ solves (\ref{thirdmu}) however it does not solve (\ref{mueq5again}) as $B$ and $C$ are necessarily non-zero. Once a solution is obtained we should be able to extract $A,B,C$ from this data.  In fact, we are able to integrate one combination of the equations. Dividing (\ref{logBeq1again}) by $B^2$ and (\ref{logCeq2again}) by $C^2$ and subtracting them  we obtain
\bea
A^4\left( \frac{1}{B^2}-\frac{1}{C^2}\right) & = & \tilde k \label{constrainteqtilde}
\eea
where $\tilde k$ is an integration constant. Further using (\ref{Aeq4again}) we obtain
\bea
B^2 C^2 (C^2-B^2) & = & k \frac{(1-\taul^2\mu^2)^4}{\mu^4}\label{constrainteq}
\eea
where $k = (\tfrac{3}{2})^4 \tilde k$.
It would have been nice to use this to find an equation of second order instead of third order, but we have not managed to do so. In any case, this constraint should be useful when doing regularity and numerical analysis as it gives some exact analytic control on the analysis. In particular, let us return to showing that once a solution for the third order equation is found, the complete solution can be reconstructed. 

 A solution $\mu$ of the third order equation  depends generically on three integration constants. Given this, $A$ can be integrated from (\ref{Aeq3again}), and contains another integration constant. We can then determine $B$ and $C$ by combining  (\ref{mueq5again}) with (\ref{constrainteqtilde}), where we regard $A$, $\mu$ and $\dot \mu$ as known functions and solve for $B$ and $C$. We have
  \bea
    B^2 & = & - \frac{2k}{9}\frac{(1-\taul^2\mu^2)^2}{\mu^2}\frac{1}{A^2}- \frac{3}{2}\frac{\taul \dot \mu }{\mu} A^2~,\nonumber\\
  C^2 & = & \frac{2k}{9}\frac{(1-\taul^2\mu^2)^2}{\mu^2}\frac{1}{A^2}- \frac{3}{2}\frac{\taul \dot \mu }{\mu} A^2~.
  \eea
Notice these are algebraic equations, so no new integration constants are introduced, and we correctly have four integration constants, one for each function.

The remaining $\theta$ dependent part of the system  leads to the following equations
\bea
\cos 2 \theta \, \partial_{\taul}\Phi + \sin 2 \theta \, \me^{\Phi} \partial_\taul C^{(0)} &=&\frac{1-\taul^{2}\mu^{2}} {2\mu\taul A} \left(\frac{C}{B}-\frac{B}{C}\right)\label{RePtaueq}~,\\
\cos 2 \theta \, \me^{\Phi} \partial_\taul C^{(0)}   -  \sin 2 \theta \, \partial_\taul \Phi& =& 0 \label{ImPtaueq}~,\\
\partial_\taul \theta &=& -\frac{1}{2} \me^{\Phi} \partial_\taul C^{(0)}\label{thetadoteq}~.
\eea
Interestingly, this decoupled set of equations can be completely integrated (assuming $\theta \neq 0$), namely we have
\bea
\partial_\taul \log \cot 2 \theta & = &  \frac{1-\taul^{2}\mu^{2}} {2\mu\taul A} \left(\frac{C}{B}-\frac{B}{C}\right)\label{integtheta}~,\\
\me^{-(\Phi-\Phi_0)} & = & \sin 2\theta~,\\
C^{(0)} & = & - \me^{\Phi_0} \cos 2\theta + C^{(0)} _0~,
\eea
where $\Phi_0$ and $C^{(0)}_0$ are two integration constants.

We have the third order equation, or equivalently a coupled system of first order equations. Once a solution is found, the phase $\theta$ can be determined by integrating (\ref{integtheta}), and finally the dilaton and axion are determined algebraically in terms of $\theta$.

Note that, for the purposes of studying (numerically) a system of first order equations, it may be convenient to consider the functions $\mu$, $A$, and then to pick one, say $B$. 
$C$ is then determined algebraically, and the $\partial_\taul C$ equation is then implied. This system reads
\bea
\dot  A&=&- \frac{2 \taul \mu^{2} }{3(1-\taul^{2}\mu^{2})} A~,\\
\dot  B &=&    \left( \frac{1+\taul^2\mu^2 }{3\taul (1-\taul^2\mu^2)} \frac{B^2}{A^2}- \frac{ 4\mu^{2} \taul}{3(1-\taul^{2}\mu^{2})} \right) B ~,\\
\dot \mu &=&          \frac{1}{3\taul}\left(\frac{B^2}{A^2} + \frac{9}{4} \frac{(1-\taul^2\mu^2)^2}{\mu^2B^2} \right)\mu ~.
\eea
Finding solutions to this ansatz is dependent on solving the third order non-linear differential equation (\ref{thirdmu}). Our preliminary studies were inconclusive and we leave the numerical study of (\ref{thirdmu}) as an open problem.

%%%%%%%%%%%%%%%%%%%%%%%%%%%%%%%%%%%%%%%%%%%%%%%%%

\section{The solution of \cite{Macpherson:2014eza}}\label{NATDsection}

 Part of the motivation for completing this work was to clarify the geometry underlying the two supersymmetric solutions in \cite{Macpherson:2014eza} which circumvented the classification of \cite{Gauntlett:2005ww}. In this final section we show that the  supersymmetric NATD-T dual of the AdS$_{5}\times$ T$^{(1,1)}$ solution in \cite{Macpherson:2014eza} satisfies our classification. We were unable to directly solve the equations of the classification to recover the solution (due to the complexity of the equations), as was done in \cite{Gauntlett:2005ww} for the Pilch-Warner solution. We instead bypassed this problem by finding the Killing spinors from which we constructed the geometry by way of the spinor bilinears. We first begin this section by writing down the solution found in \cite{Macpherson:2014eza}. 

 We use the coordinates $x_{1}=\rho \sin \chi~, x_{2}=\rho \cos\chi$ and for simplicity set $\alpha'=1$. The d=10 metric in string frame\footnote{Recall that the classification is in Einstein frame.} is 
\begin{gather}
\ud s^{2}=\ud s^{2}(AdS_{5})+L^{2}\lambda_{1}^{2}\ud \theta_{1}^{2} +\frac{1}{L^{2} P Q}\left((L^{4}\lambda^{2}\lambda_{1}^{2}+x_{1}^{2})\ud x_{1}+x_{1}x_{2}\ud x_{2}\right)^{2}\nonumber\\
+\frac{L^{2} \lambda_{1}^{2}}{P}\ud x_{2}^{2}
+ \frac{1}{L^2 W Q}(Q \ud \phi_{1}-\lambda^{2} x_{1} x_{2}\cos \theta_{1}  \ud x_{1}-\lambda^{2}(L^{4}\lambda_{1}^{4}+x_{2}^{2})\cos\theta_{1}\ud x_{2})^{2}\nonumber\\
+\frac{L^{2} \lambda^{2}\lambda_{1}^{4}x_{1}^{2}\sin^{2}\theta_{1}}{W}\ud \xi^{2}~,\label{NATDmetric}
\end{gather}
where
\begin{gather*}
Q=L^{4}\lambda^{2}\lambda_{1}^{4}+\lambda_{1}^{2} x_{1}^{2}+\lambda^{2} x_{2}^{2}~,~~ W=\lambda_{1}^{2}Q \sin^{2}\theta_{1} +\lambda^{2}\lambda_{1}^{2} x_{1}^{2}\cos^{2}\theta_{1}~,~~
P=L^{4}\lambda^{2}\lambda_{1}^{2}+x_{1}^{2}~.
\end{gather*}
The constants $\lambda$ and $\lambda_{1}$ take the values $1/3$ and $1/\sqrt{6}$ respectively and $L$ is the radius of AdS$_{5}$. The dilaton is 
\begin{equation}
\me^{-2 \Phi}=L^{4} W~,
\end{equation}
whilst the NS-NS two-form is given by\footnote{We correct a minor typographical error here by adding the $\cos\theta_{1}$ term in front of $\ud \phi_{1}$.}
\begin{gather}
B=\left.\left.-\frac{\lambda_{1}^{2}x_{1}}{W}\right(\lambda^{2}x_{1}\cos\theta_{1}  \ud \phi_{1}+\lambda^{2}x_{2}\sin^{2}\theta_{1} \ud x_{1}-x_{1}(\lambda^{2}\cos^{2}\theta_{1}+\lambda_{1}^{2}\sin^{2}\theta_{1}) \ud x_{2}\right)\wedge \ud \xi~.
\end{gather}
The non-zero RR-fluxes\footnote{These are the ones that appear in the equations of motion, $F_{n}=\ud C_{n-1}-C_{n-3}\wedge \ud B$.} are
\bea
F_{1}&=&4 L^{4}\lambda \lambda_{1}^{4}\sin \theta_{1}\ud \theta_{1}~,\\
F_{3}&=&\frac{4 L^{4} \lambda \lambda_{1}^{6}x_{1} \sin\theta_{1}}{W}\ud \theta_{1} \wedge \ud \xi\wedge \nonumber\\%%
&&\left[ \lambda^{2} x_{2}\sin^{2}\theta_{1} \ud x_{1}-x_{1}(\lambda^{2}\cos^{2}\theta_{1}+\lambda_{1}^{2}\sin^{2}\theta_{1})\ud x_{2}+\lambda^{2}x_{1}\cos\theta_{1}\ud \phi_{1}\right]~,
\eea
and of course their hodge duals. In the notation of this classification the corresponding elements are
\bea
m&=&\frac{1}{L}\label{Natdm}~,\\
\taul&=&L^{2} \lambda_{1}^{2} x_{1} \sin\theta_{1}~,\label{Natdtau}\\
\mu&=&\frac{1}{L^{2}\sqrt{W}}=\me^{\Phi}~,\\
\ud \psi&=&-\ud \xi~,\\
\beta&=&(-x_{1}\cos \theta_{1} \ud x_{1}-x_{2} \cos \theta_{1} \ud x_{2}+L^{4}\lambda_{1}^{4}\sin \theta_{1} \ud \theta_{1}+x_{2} \ud \phi_{1})~,\\
\sigma&=&\frac{L\lambda_{1}^{2}}{W^{1/4}}[x_{1}x_{2}\sin^{2}\theta_{1}\ud x_{1}+(L^{4}\lambda_{1}^{4}+x_{2}^{2})\sin^{2}\theta_{1}\ud x_{2}+x_{1}^{2}\cos \theta_{1} \ud \phi_{1}\nonumber\\%%%%%%%%
&&+\ii L^{2}\sqrt{W}(\cos\theta_{1} \ud x_{2}+x_{2}\sin \theta_{1} \ud \theta_{1}-\ud \phi_{1})]~.\label{Natdsigma}
\eea
Further details on the derivation of this dictionary is presented in appendix \ref{NATDapp}. One may check that (\ref{NATDmetric}) takes the form of (\ref{Classmetric}) with these identifications.
For the explicit form of the NS-NS two form we find
\begin{gather}
B=\frac{\me^{\Phi/2}\mu}{9m^{2}}\Real[\sigma]\wedge \ud \psi-\ud x_{2}\wedge \ud \psi~,
\end{gather}
whilst $C^{(2)}$ is not given in \cite{Macpherson:2014eza} for us to compare with, however it is trivial to show that $F_{3}$ agrees with that derived from the general expressions (\ref{Beq}) and  (\ref{C2}).

 We have checked that this solution satisfies all the conditions of the classification, as an illustrative example we present the solution of (\ref{beta}). First define the function $E=(L^{4}\lambda_{1}^{4}+x_{2}^{2})\sin^{2}\theta_{1}+x_{1}^{2}\cos^{2}\theta_{1}$. A short calculation gives
\begin{gather}
\ii \sigma^{*}\wedge \sigma-2\taul \ud \taul \wedge \beta=2 L^{16}\lambda_{1}^{4}E[\ud x_{2}\wedge \ud \phi_{1}+x_{1}\sin\theta_{1} \ud \theta_{1} \wedge \ud x_{1}+x_{2}\sin\theta_{1} \ud \theta_{1} \wedge \ud x_{2}]~,
\end{gather}
whilst
\begin{gather}
\frac{3(1-\taul^{2}\mu^{2}) }{\mu^{2}} \ud \beta= L^{16} \lambda \lambda_{1}^{2} E [ \ud x_{2} \wedge \ud \phi_{1}+x_{1}\sin\theta_{1} \ud \theta_{1} \wedge \ud x_{1}+x_{2}\sin\theta_{1} \ud \theta_{1} \wedge \ud x_{2}]~.
\end{gather}
Upon substituting the values of the constants, $\lambda$ and $\lambda_{1}$ we find that they are equal. The equation for $\sigma$ follows similarly but is vastly more complicated than the one illustrated above and for this reason we do not present it.

 In section \ref{SectionLocalCoords} we saw that the integrable almost product structure implied that the one-forms $\beta$ and $\sigma$ had no $\ud \tau$ term, we would like to verify this. To do so we must write the one-forms in the form (\ref{3coords}). To this end, we make the change of coordinates
\bea
x&=&\phi_{1}~,\label{changeofcoords1}\\
y_{1}&=&\frac{1}{2}(x_{1}^{2}+x_{2}^{2})+L^{4}\lambda_{1}^{4} \ln( \cos \theta_{1})~,\\
y_{2}&=&\ln\left( \frac{x_{2}}{\cos \theta_{1}}\right)~,\\
\taul&=&L^{2}\lambda_{1}^{2} x_{1} \sin \theta_{1}~.\label{changeofcoords4}
\eea
In these coordinates the coefficients for the one-forms, in the notation of (\ref{3coords}), are 
\bea
\gamma_{x}&=&  x_{2}~,\qquad ~~~~~~~~~~\gamma_{y_{1}}=-\cos \theta_{1}~,\qquad~~~\gamma_{y_{2}}=0~,\\
 \rho_{x}&=&\frac{L\lambda_{1}^{2}x_{1}^{2}\cos \theta_{1}}{W^{1/4}}~, ~~~  \rho_{y_{1}} =\frac{L\lambda_{1}^{2}x_{2} \sin^{2}\theta_{1}}{W^{1/4}}~,~~ \rho_{ y_{2}}= \frac{L^5\lambda_{1}^{6}x_{2} \sin^{2}\theta_{1}}{W^{1/4}}~,\\
 \kappa_{ x}&=&-L^{3}\lambda_{1}^2 W^{1/4}~,\quad~ \kappa_{y_{1}}=0~,~~~~~~~~~~~~~~~~~~ \kappa_{y_{2}}=L^{3}\lambda_{1}^{2}  x_{2}\cos\theta_{1} W^{1/4}~.
\eea
It is clear that this satisfies the integrable almost product structure. We have again checked that with these new coordinates the equations of the classification are satisfied and once again the equations to solve are very complicated. We had hoped this solution would have motivated further ansatz, unfortunately this was not the case. Interestingly this solution has an additional Killing vector, $\partial_{x}$, to what the classification implies. Imposing this extra Killing direction does not give much in the way of simplification of the equations and so this ansatz was swiftly dropped in favour of the ones we have presented.

 We note that this solution, like our one, is singular \cite{Macpherson:2014eza}. The Ricci tensor blows up as $\theta_{1}\rightarrow 0$ or $\pi$ whilst $x_{1}\rightarrow 0$. Furthermore the dilaton also blows up at these points. Computing the invariants $R_{\mu\nu}R^{\mu\nu}$ and $R_{\mu_{1}..\mu_{4}}R^{\mu_{1}..\mu_{4}}$ we also find that these are singular at these points but only these points. This solution therefore exhibits two singular points.

 Though the solution is singular it would still be interesting to interpret this solution's field theory dual and also its brane realisation. A method was proposed in \cite{Lozano:2016kum} where they considered the type IIA non-Abelian T dual of AdS$_{5}\times S^{5}$ and propose a  a D4/NS5 brane set-up and a linear quiver to describe its dual SCFT. 

In \cite{Macpherson:2014eza} they also present another supersymmetric type IIB solution with $F_{5}=0$, namely the NATD-T dual of the AdS$_{5}\times Y^{p,q}$ solution. This solution will also satisfy the classification presented here however we have not checked the details.

%%%%%%%%%%%%%%%%%%%%%%%%%%%%%%%%%%%%%%%
\section{Conclusions}\label{Conclusion}

This work has plugged the remaining gap in the classification of all AdS$_{5}$ supersymmetric solutions of type IIB supergravity. Together with \cite{{Gauntlett:2005ww},{Gauntlett:2004zh},{Apruzzi:2015zna}} our work concludes the classification of all supersymmetric AdS$_{5}$ solutions of $d=10$ and $d=11$ supergravity. We find that the geometry of $M_{5}$ is different to that of the $F_{5}\not=0$ case. It should be possible to interpret these results in terms of the ``Exceptional Sasaki-Einstein (ESE) geometry'' of \cite{Ashmore:2016qvs}\footnote{We thank Daniel Waldram for clarifications on this point}. It would be interesting to see how the ESE structure is interpreted in terms of the bilinears. A similar analysis was carried out in \cite{Ashmore:2016qvs} for the case of $F_{5}\not=0$.

One of the motivations for doing this work was to find new non-singular supersymmetric solutions relevant for AdS/CFT. From \cite{Macpherson:2014eza} we knew that singular supersymmetric solutions did exist, however the only solution we found was once again singular. In particular from the analysis performed in section \ref{secP=0} and appendix \ref{Appanalysis} we conclude that there are no non-singular solutions with $P=0$. Contrast this with the $F_{5}\not=0$ case \cite{Gauntlett:2005ww} where one finds the infinitely many Sasaki-Einstein solutions and the Pilch-Warner solution (which has $P=0$), whilst in type IIA \cite{Apruzzi:2015zna} one finds infinitely many massive IIA solutions and recovers previously known massless solutions such as the Maldacena-N\'u\~nez solution. Moreover in eleven dimensions many new solutions were found \cite{Gauntlett:2004zh}. It is therefore disappointing that we have been unable to find new non-singular solutions. 

However there are solution generating techniques one may use to find new solutions with $F_{5}=0$ (and also $F_{5}\not= 0$). As pointed out in \cite{Macpherson:2014eza} if one begins with a Sasaki-Einstein solution with at least $SU(2)\times U(1)\times U(1)$ and follows their procedure for applying the Non-Abelian T-duality followed by the T-duality one obtains solutions with $F_{5}=0$, whether they are supersymmetric and non-singular is case dependent. Moreover one may obtain solutions with $F_{5}=0$ by T-dualising a IIA solution whose $F_{4}$ flux has a leg over the direction that is being dualised over for all components, once again supersymmetry and regularity is case dependent.

An interesting class of solutions are those which can be represented in both IIA, IIB and possibly also in eleven-dimensional supergravity. It may be fruitful to compare the supersymmetry conditions of this classification with the different cases, \cite{Gauntlett:2004zh} and \cite{Apruzzi:2015zna}. More concretely if we assume $\partial_{x}$ is a Killing vector we may T-dualise over it to type IIA where we are then able to compare this classification with \cite{Apruzzi:2015zna}. Uplifting to 11d allows us to compare with \cite{Gauntlett:2004zh}.

%%%%%%%%%%%%%%%%%%%%%%%%%%%%%%%%%%%%%%%%%%%%%%%%%%%%%%%%%%%%%%%%%%%%%%%
\subsection*{Acknowledgments}
\noindent

I would like to thank Dario Martelli for suggesting the research topic and for guidance throughout the completion of the work.  I would also like to thank the authors of \cite{Macpherson:2014eza} for helpful clarifications on their paper. Thanks also go to Eoin \'O Colg\'ain and Daniel Waldram for comments on an earlier version of this work. My work is supported by an STFC studentship, number ST/N504361/1.

%%%%%%%%%%%%%%%%%%%%%%%%%%%%%%%%%%%%%%%%%%%%%%%%%%%%%%%%%%%%%%%%%%%%%%%

\appendix

\section{Bilinear definitions and the orthonormal frame}\label{Oframe}

We define all the bilinears appearing in the paper. The scalar bilinears are
\bea
A &\equiv &\frac{1}{2}(\bar{\xi}_{1}\xi_{1}+\bar{\xi}_{2}\xi_{2})~,\nonumber\\
A\sin\zeta &\equiv&\frac{1}{2}(\bar{\xi}_{1}\xi_{1}-\bar{\xi}_{2}\xi_{2})~,\nonumber\\
S &\equiv&\bar{\xi}^{c}_{2}\xi_{1}\nonumber~,\\
Z &\equiv&\bar{\xi}_{2}\xi_{1}\label{scalars}~.
\eea
The vector bilinears are
\bea
K^{m}  &\equiv & \bar{\xi}_{1}^{c}\gamma^{m}\xi_{2}~,\nonumber\\
K_{3}^{m} &\equiv & \bar{\xi}_{2}\gamma^{m}\xi_{1}~,\nonumber\\
K_{4}^{m} &\equiv & \frac{1}{2}(\bar{\xi}_{1}\gamma^{m}\xi_{1}-\bar{\xi}_{2}\gamma^{m}\xi_{2})~,\nonumber\\
K_{5}^{m} &\equiv & \frac{1}{2}(\bar{\xi}_{1}\gamma^{m}\xi_{1}+\bar{\xi}_{2}\gamma^{m}\xi_{2})~.\label{vectors}
\eea
The two-form bilinears are
\bea
W_{mn}&\equiv &-\bar{\xi}_{2}\gamma_{mn}\xi_{1}~,\nonumber\\
V_{mn}&\equiv &-\frac{\ii}{2}(\bar{\xi}_{1}\gamma_{mn}\xi_{1}-\bar{\xi}_{2}\gamma_{mn}\xi_{2})~,\nonumber\\
U_{mn}&\equiv &-\frac{\ii}{2}(\bar{\xi}_{1}\gamma_{mn}\xi_{1}+\bar{\xi}_{2}\gamma_{mn}\xi_{2})~,   \\
X_{mn}&\equiv & \bar{\xi}_{1}^{c}\gamma_{mn}\xi_{1}~,\nonumber\\
Y_{mn}&\equiv &\bar{\xi}_{2}^{c}\gamma_{mn}\xi_{2}~,\nonumber
\eea
One finds that they satisfy the following algebraic relations
\bea
K_{5}&=&\sin \zeta ~K_{4}+\Real[Z^{*}K_{3}]-\Real[S^{*} K]  ~, \label{K5alg}\\
0&=&\sin\zeta V-U-\frac{\ii}{2} K^{*}\wedge K+\Real[\ii Z^{*}W]~,\\
S^{*}X&=&(1+\sin\zeta)W-(K_{4}+K_{5})\wedge K_{3}~,\\
S^{*}Y&=&(1-\sin\zeta)W^{*}-(K_{4}-K_{5})\wedge K_{3}^{*}~.
\eea

These relations may be computed by making use of Fierz identities, however we find it simpler to compute these by using an orthonormal frame which we shall construct below. Following \cite{Gauntlett:2005ww} we take the basis of gamma matrices of Cliff$(5)$ to be
 \bea
\gamma^{1}&=&\begin{pmatrix}1&0\\0&-1\end{pmatrix}\otimes I\nonumber\\
 \gamma^{2}&=&\begin{pmatrix} 0&1\\1&0\end{pmatrix}\otimes I \nonumber\\
 \gamma^{a}&=&\begin{pmatrix} 0&-1\\1&0\end{pmatrix}\otimes \tau^{a}
 \eea
 where $\tau^{a}=-\ii \sigma^{a}$ and $\sigma^{a}$ are the Pauli matrices. In this basis the charge conjugation intertwiner is given by $C=I\otimes\tau^{2}$. we label the corresponding basis by $e^{i}$. We decompose the spinors $\xi_{i}$ as $s_{i}\otimes \theta_{i}$ where $s_{i}$ are spinors of Cliff(3) and $\theta_{i}$ spinors of Cliff(2). At the moment the basis is completely arbitrary which allows us to impose that the two vectors $K_{4}$ and $K_{5}$ lie in the $(e^{1}-e^{2})$ plane and in particular $K_{5}$ to be parallel with $e^{1}$. We find
 \begin{equation}
s_{1}=\sqrt{2}\begin{pmatrix}\cos\theta\cos\phi\\-\sin\theta\sin\phi\end{pmatrix}~, \quad s_{2}=\sqrt{2}\begin{pmatrix}\sin \theta\cos\phi\\\cos\theta\sin\phi\end{pmatrix}
 \end{equation}
 where we have set $\bar{\theta}_{i}\theta_{i}=1$ and added suitable normalization to enforce $A=1$. We can now write the scalar and vector bilinears as functions of $\theta,\phi, \theta_{i}$. Requiring $\sin \zeta=0$ implies that $\cos 2\theta=0$ otherwise $\cos 2\phi=0$ which then implies $K_{5}=0$. Choosing $K_{3}$ to lie in the $(e^{3}\text{-}e^{4})$ plane one can choose:
 \begin{equation}
 \theta_{1}=\begin{pmatrix} e^{\ii \alpha}\\0\end{pmatrix},\quad \theta_{2}=\begin{pmatrix}0\\ e^{\ii \alpha}\end{pmatrix}
 \end{equation}
 from which we obtain the final form of the vector bilinears
 \begin{gather}
 K_{5}=\cos 2\phi e^{1}~,\quad K_{4}=-\sin 2\phi e^{2}~,\quad K_{3}=\sin 2\phi (e^{4}-\ii e^{3})~, \nonumber\\
 K=\me^{2\ii\alpha}e^{1}-\ii \sin 2 \phi e^{2\ii \alpha} e^{5}~,
 \end{gather}
 and the one non-trivial scalar bilinear
 \begin{equation}
 S=-\me^{2\ii \alpha}\cos 2\phi~.\label{S}
 \end{equation}
The two-forms in terms of this orthonormal basis are
\begin{gather}
U=-\sin2\phi e^{15}~,\quad V=e^{34}-\cos 2\phi e^{25}~,\quad W=(\ii \cos 2 \phi e^{5}-e^{2})\wedge(e^{4}-\ii e^{3})~,\nonumber\\
X=\me^{2 \ii \alpha}(\sin 2\phi e^{1}+\cos 2 \phi e^{2}- \ii e^{5})\wedge (e^{4}-\ii e^{3})~,\quad\nonumber\\
 Y=\me^{2 \ii \alpha}(-\sin 2\phi e^{1}+\cos 2\phi e^{2}+\ii e^{5})\wedge(e^{4}+\ii e^{3})~.
\end{gather}

\section{Algebraic analysis of (\ref{secordermu}) for $c\not=0$}\label{Appanalysis}

For $c\not=0$ equations (\ref{hatbeta})-(\ref{hatI}) have solution:
\bea
\hat{\beta}&=&\frac{1}{c}\tau_{3}~,\\
\hat{R}&=&\sqrt{\frac{3}{2|c|}}\tau_{2}~,\\
\hat{I}&=& \sqrt{\frac{3}{2|c|}}\tau_{1}~,\label{Riccic<0}
\eea
where $\tau_{i}$ are the $SU(2)$ left invariant one-forms if $c>0$ and the $SL(2,\mathbb{R})$ left invariant one-forms if $c<0$ \footnote{ The left invariant $SU(2)$ one-forms satisfy $\ud \tau_{1}=\tau_{2}\wedge \tau_{3}$ and cyclic permutations, whilst the $SL(2,\mathbb{R})$ one-forms satisfy $\ud \tau_{1}=\tau_{2}\wedge \tau_{3},~\ud \tau_{2}=\tau_{3}\wedge \tau_{1},~\ud \tau_{3}=\tau_{2}\wedge \tau_{1}$.}.
The metric becomes
\begin{gather}
9 m^{2}\ud s^{2}=\taul^{2} \mu^{2} \ud \psi^{2}+\frac{\mu^{2}}{1-\taul^{2}\mu^{2}}\ud \taul^{2}
+\mu^{2}(1-\taul^{2}\mu^{2})\left(\frac{1}{\taul^{2}\dot{\mu}^{2}}\tau_{3}^{2}+\frac{3}{2 |\taul \dot{\mu}|\mu}(\tau_{1}^{2}+\tau_{2}^{2})\right)\label{metriccnot0}
\end{gather}
We have managed to find a solution to the differential equation (\ref{secordermu}) when $c\not=0$, namely $\mu=\sqrt{3}/\taul$. Unfortunately this is not an admissible solution as it gives a metric with the wrong signature which can be clearly seen from the above.

We now present some algebraic analysis on the existence of regular solutions to (\ref{secordermu}), considering first the case $c>0$ and then the case $c<0$. We must find the range of the coordinate $\taul$ and show that the metric is regular for all values of $\taul$ in this range. To do so we find values of $\taul$ for which the metric shrinks, equivalently some function of the metric becomes zero, yet the metric remains non-singular. Upon using (\ref{secordermu}) and its first derivative in $\taul$, we find that the Ricci scalar is given in the two cases by
\bea
R_{c>0}&=& \frac{m^{2}}{2 \mu^{4}}(56 \mu^{4}+240\dot{\mu}  \taul \mu^{3}+\dot{\mu}^{2}(9+171 \taul^{2}\mu^{2}))~,\\
R_{c<0}&=&R_{c>0}+\frac{24 m^{2} \taul \dot{\mu}}{\mu (1-\taul^{2}\mu^{2})}~.
\eea

\subsection*{$c>0$ analysis}
We first consider the case where the function $1-\taul^{2}\mu^{2}$ vanishes, let this point be $\taul_{0}$. Near to $\taul_{0}$ we may write
\bea
1-\taul^{2}\mu^{2}\simeq \bet (\taul-\taul_{0})^{2\alpha}
\eea 
for some constants $\alpha$ and $\bet$. Making the change of coordinate\footnote{Note that we have implicitly assumed $\alpha\not=1$ here. However for $\alpha=1$ one finds that the Ricci-scalar has a singularity as $\taul\rightarrow\taul_{0}$.},
\bea
r=\frac{(\taul-\taul_{0})^{1-\alpha}}{\bet^(1/2) (1-\alpha)}
\eea
we have 
\bea 
\frac{\ud \taul^{2}}{1-\taul^{2}\mu^{2}}=\ud r^{2}~,
\eea
and
\bea
1-(\taul \mu)^{2}= \bet^{\frac{1}{1-\alpha}}((1-\alpha)r)^{\frac{2 \alpha}{1-\alpha}}~.
\eea
Requiring that the latter expression is proportional to $r^{2}$, as it should be for a regular solution, we find $\alpha=1/2$. Near to $\taul_{0}$ the metric takes the form
\bea
9 m^{2}\ud s^{2}=\mu^{2}\left[ \taul^{2}\ud \psi^{2}+\ud r^{2}+\frac{r^{2}\bet^{2}}{4}\lb   \frac{1}{\taul^{2}\dot{\mu}^{2}}\tau_{3}^{2}+\frac{3}{2 |\taul  \dot{\mu}| \mu}(\tau_{1}^{2}+\tau_{2}^{2})                          \rb    \right]~. \label{metricanalysis}
\eea
For regularity we require that the metric looks locally like $S^{1}\times \mathbb{R}^{4}$. For this to occur we require the factors in front of the left invariant one-forms to be equal and the overall factor to be $r^{2}/4$. Using the expression for $\mu$ near $\taul_{0}$ we find $\taul_{0}\leq0$ and $\bet=-\frac{2}{3 \taul_{0}}$, we have implicitly assumed that we are away from $\taul_{0}=0$ to obtain $\bet$. Notice however that if we are at $\taul=0$ then the solution will not be regular as $\mu$ is then necessarily unbounded in order to satisfy $1-\taul_{0}^{2}\mu^{2}=0$. We find that for any $\taul_{0}$ strictly negative with $\mu$ satisfying $1-\taul^{2} \mu^{2}=0$ at $\taul_{0}$ this will define an endpoint of the range of $\taul$ and the metric will be regular at this point. 

We may ask whether it is possible for there to be two such values of $\taul$, for which $1-\taul^{2}\mu^{2}=0$ away from $\taul=0$. Assume that $\taul_{1}$ and $\taul_{2}$ are two such values, and that there is no point $\taul_{3}\in (\taul_{1},\taul_{2})$ such that $1-\taul_{3}^{2}\mu(\taul_{3})^{2}=0$, otherwise we have not chosen our range for $\taul$ correctly. Without loss of generality and with the previous analysis in mind set $\taul_{1}<\taul_{2}<0$. Near to $\taul_{a}$, $a=1,2$, we have $\dot{\mu}(\taul)=-2/(3 \taul^{2})$. Therefore for $\taul= \taul_{1}+\epsilon_{1}$, with $\epsilon_{1}$ a small positive number,  $\mu(\taul_{1}+\epsilon_{1})<-1/(\taul_{1}+\epsilon_{1})$ however near to $\taul_{2}$ we have, for $\epsilon_{2}$ a small positive number, $\mu(\taul_{2}-\epsilon_{2})>-1/(\taul_{2}-\epsilon_{2})$. With the additional and not unreasonable assumption that $\mu$ is continuous we must have that at some point $\taul_{3}\in(\taul_{1},\taul_{2})$ that $1-(\taul_{3}\mu(\taul_{3}))^{2}=0$ and hence we reach a contradiction as we assumed no $\taul_{3}$ existed. We conclude that no two such points exist.

Assume now that $\taul=0$ is a regular boundary solution. For regularity it is necessary that $\mu$ takes a finite value at $\taul=0$ or that it diverges as $ O(1/\taul)$. A regular solution occurs if the last bracketed term in (\ref{metriccnot0}) is finite in the limit as $\taul$ goes to 0 or it goes to zero as $\taul^{2}$ and has the metric of a three-sphere. If we consider these cases then $\mu\propto \log\taul$ or $\mu\propto \taul^{-\alpha},~\alpha>1$ as $\taul\rightarrow 0$. However one now finds that the full $d=10$ metric has singular Ricci scalar as $\taul\rightarrow 0$ in both cases. Moreover if we expand (\ref{secordermu}) about $\tau=0$ we find that the only solution with this asymptotic behaviour is the true solution that gives the incorrect signature. This suggests that $\taul=0$ is not a boundary condition that gives a non-singular metric.

The remaining possibilities are $\mu(\taul_{0})=0$ for some $\taul_{0}$, that $\dot{\mu}(\taul_{0})=0$ or that $\taul\rightarrow -\infty$. We first look at the $\dot{\mu}(\taul_{0})=0$ case. Equation (\ref{secordermu}) implies that either $\taul_{0}=0$, $1-\taul_{0}^{2}\mu^{2}(\taul_{0})=0$ or $\ddot{\mu}(\taul_{0})=0$ at $\taul_{0}$. We can rule out both the first and second choices from our previous analysis, leaving us to conclude that $\ddot{\mu}(\taul_{0})=0$. We then find that all the derivatives of $\mu$ vanish at this point by taking further derivatives of (\ref{secordermu}) and evaluating at $\taul_{0}$. Assuming, not unreasonably, that $\mu$ is analytic at this point we conclude that $\mu$ is a constant everywhere violating $c\not=0$.

We next consider the possibility that $\mu(\taul_{0})=0$. Then, near to $\taul_{0}$, we may write
\bea
\mu=\bet (\taul-\taul_{0})^{ \alpha}~,
\eea
with $\alpha>0$ and the metric takes the form
\bea
9 m^{2}\ud s^{2}=\mu^{2}\left[ \taul^{2}\ud \psi^{2}+\ud \taul^{2}+\frac{1}{\alpha^{2}\beta^{2}\taul_{0}^{2}(\taul-\taul_{0})^{2(\alpha-1)}}\tau_{3}^{2}+\frac{3}{2 \alpha \beta^{2}\taul_{0} (\taul-\taul_{0})^{2 \alpha-1}}(\tau_{1}^{2}+\tau_{2}^{2})\right]~.
\eea
One can see immediately that this is not regular for any $\alpha>0$ and $\taul_{0}$ as the Ricci scalar diverges. For $\mu$ diverging at $\taul_{0}$ one still requires $1-\taul^{2}\mu^{2}>0$, for a metric with the correct signature, and therefore $\taul_{0}=0$ which was covered in a previous case. 

Finally we study the possibility that $\taul\rightarrow -\infty$. It is best if we make the change of coordinate $\taul=-1/r$. With this change of coordinate the metric takes the form
\bea
9m^{2}\ud s^{2}=\mu^{2}\left[\frac{1}{r^{2}}\ud \psi^{2}+\frac{r^{4}}{1-\frac{\mu^{2}}{r^{2}}}\ud r^{2}+\frac{r^{2}-\mu^{2}}{r^{2}}\lb \frac{1}{r^{2}\mu'^{2}}\tau_{3}^{2}+\frac{3}{2 r \mu' \mu}(\tau_{1}^{2}+\tau_{2}^{2})\rb\right]
\eea
We still require that $1-\mu^{2}/r^{2}>0$ and so for small $r$, $\mu$ must take the form 
\bea
\mu=a_{1}r+a_{2}r^{2}+...~,
\eea
with $|a_{1}|<1$. From looking at the last term in (\ref{metriccnot0}) we see that we need $\mu=a r$. With a further coordinate transformation $s=r^{4}/4$ the metric takes the form
\bea
9m^{2}\ud s^{2}=a^2\lb \ud \psi^{2}+\frac{\ud s^{2}}{1-a^{2}}+\frac{1-a^{2}}{a^{2}}\lb\tau_{3}^{2}+\frac{3}{2}(\tau_{1}^{2}+\tau_{2}^{2})\rb\rb~.
\eea
The metric takes the form of $S^{1}\times \mathbb{R}\times S^{3}$ where the $S^{3}$  is squashed. Note however as $r\rightarrow 0$ we have the form of $\mu$ in this limit and inserting this into (\ref{secordermu}) we find that $a=\sqrt{3}$ and hence the metric has the wrong signature. This suggests that there are no non-singular solutions for $c>0$ and we turn our attention to $c<0$ in the following subsection.

\subsection*{$c<0$ analysis}

We now consider the case of $c<0$, recall that now $\tau_{i}$ are the left invariant $SL(2,\mathbb{R})$ one-forms. Most of the arguments from the $c>0$ case are still applicable and we shall make use of these when possible. Note that the possibility of $1-\taul^{2}\mu^{2}=0$ at $\taul_{0}$ will no longer give a non-singular metric as before.This can be seen directly from the Ricci scalar in equation \ref{Riccic<0}.

Assume that $\taul=0$ is a boundary condition. In the previous argument for $\taul=0$ in the $c>0$ case, we did not reference the particular form of the metric until computing the Ricci scalar of the full $d=10$ metric, once again this diverges as $\taul\rightarrow 0$ and this suggests that $\taul=0$ is not a regular boundary condition. The argument that forbid non-singular solutions with $\dot{\mu}(\taul_{0})=0$ still applies in the $c<0$ case and so this is also not possible. Moreover we cannot have $\mu(\taul_{0})=0$ for the same reasons as in the $c<0$ case as the Ricci scalar diverges. Note that the final possibility for a boundary value is $\taul\rightarrow \pm \infty$. As $\taul=0$ gives a singular point for the manifold we cannot have the range to be $\taul\in (-\infty,\infty)$ and therefore there are no two points for $\taul$ to take a value in. If one completes the analysis for $\taul\rightarrow \pm \infty$ one again finds that the manifold is singular at these points.

From the analysis of this and the previous subsection we conclude that no non-singular analytic solutions with $P=0$ exist with $c>0$ and $c<0$.

%%%%%%%%%%%%%%%%%%%%%

\section{More details on the solution of \cite{Macpherson:2014eza}} \label{NATDapp}

In this appendix we present details about the derivation of (\ref{Natdm})-(\ref{Natdsigma}). We make no claims that all the work in this appendix is original, only the final expressions (\ref{Natdm})-(\ref{Natdsigma}). As pointed out in the text we were unable to solve the equations of the classification in order to recover this solution, in hindsight this was to be expected as it solves very non-trivial equations compared to the ansatz we have considered. Instead we found the Killing spinor of the NATD-T solution and from it constructed the spinor bilinears which allowed us to recover the solution. One may solve the Killing spinor equations directly for the NATD-T solution however this is very difficult and may be avoided. Instead one can use the Killing spinors of $T^{(1,1)}$, which are relatively simple to find, and transform them under the corresponding NATD and T dualities. It is this method that we present below.

The Buscher rules \cite{Buscher:1987sk} give the transformation of the NS-NS sector under T-duality whilst \cite{ Bergshoeff:1995as} first gave the transformation of the RR-fluxes. The transformation of the Killing spinors was found in \cite{Hassan:1999bv}.
It is also well known how the geometry changes under NATD, see \cite{delaOssa:1992vci} for the transformation of the NS-NS sector, though we shall follow the conventions in \cite{Itsios:2013wd}. The transformation of the RR-fluxes was found in \cite{Sfetsos:2010uq} whilst in \cite{Kelekci:2014ima} it was found how a Killing spinor transforms under NATD. We shall briefly present the transformation of the Killing spinors under both NATD and T-duality for the ease of the reader. 

Under a NATD or T-duality there is some ambiguity with the transformation of the vielbeins. Left and right movers of the world-sheet have different transformation properties and therefore define two different frame fields. These two frames must be equivalent as they define the same geometry and so are related by a Lorentz transformation of the form:
\begin{equation}
\hat{e}_{+}=\Lambda \hat{e}_{-}~.
\end{equation}
This Lorentz transformation induces an action on spinors by the matrix $\Omega$ which satisfies
\bea
\Omega^{-1}\Gamma^{a}\Omega=\tensor{\Lambda}{^{a}_{b}}\gamma^{b}~.
\eea
Type IIB supersymmetry is parametrised by two $d=10$ Majorana-Weyl spinors of the same chirality whilst type IIA is paramtrised by two $d=10$ Majorana-Weyl spinors of opposite chirality. We shall denote these two spinors generically as $\chi_{1}$ and $\chi_{2}$, their chiralities are unimportant for the calculation and so we do not distinguish their chiralities. Under a NATD or T-duality 
\bea
\chi_{1}\rightarrow \chi_{1}\qquad \chi_{2}\rightarrow \Omega^{-1}\chi_{2}~.\label{Ttrans}
\eea
where for a T-duality along a Killing vector, $\partial_{x}$, $\Omega$ takes the form
\bea
\Omega_{U(1)}^{-1}=-\frac{1}{\sqrt{G_{xx}}}\Gamma_{11}\Gamma_{x}~,
\eea
where $x$ is a curved index on $\Gamma_{x}$. Under a NATD, with respect to an $SU(2)$ isometry along the flat directions $1,2$ and $3$, $\Omega$ takes the form
\bea 
\Omega_{SU(2)}^{-1}=-\frac{\Gamma^{(11)}}{\sqrt{1+\zeta^{2}}}(\Gamma^{123}+\zeta_{a}\Gamma^{a})~,
\eea
where for our purposes
\begin{gather}
\zeta^{1}=\frac{x_{1} \cos\xi}{L^{2}\lambda_{1}\lambda}~,\quad \zeta^{2}=\frac{x_{1} \sin \xi}{L^{2}\lambda_{1}\lambda}~,\quad \zeta^{3}=\frac{x_{2}}{L^{2}\lambda_{1}^{2}}~.
\end{gather}
Note that both $\Omega$'s defined above are unitary in our basis.

To begin we solve the Killing spinor equations of the Klebanov-Witten solution, $T^{(1,1)}$,  in the canonical vielbein basis for performing the NATD
\bea
e^{\theta_{1}}=L \lambda_{1}\ud \theta_{1}~, ~~e^{\phi_{1}}=L\lambda_{1}\sin \theta_{1}\ud \phi_{1}~,\nonumber\\
 e^{1,2}=L \lambda_{1}\tau_{1,2}~, ~~e^{3}=L \lambda (\tau_{3}+\cos \theta_{1}\ud \phi_{1})~,
\eea
where $\tau_{i}$ are the left invaraint $SU(2)$ one-forms. With this basis, the Killing spinors are
\begin{equation}
\chi_{1}=\frac{1}{2}\begin{pmatrix} 1\\0\\ \ii \\0 \end{pmatrix},\qquad \chi_{2}=\frac{1}{2}\begin{pmatrix} \ii\\0\\-1\\ 0 \end{pmatrix}~,
\end{equation}
where the choice of normalization is for later convenience. From these two spinors we may construct $\xi_{1}$ and $\xi_{2}$ as used in the classification
\bea
\xi_{1}=\chi_{1}+\ii \chi_{2}~, \quad \xi_{2}=\chi_{1}-\ii \chi_{2}~,
\eea
note that it is the $\chi$'s that transform as (\ref{Ttrans}) and not the $\xi$'s. Under the NATD the Killing spinors become
\bea
\chi_{1}\rightarrow \chi_{1}~,\quad \chi_{2}\rightarrow \Omega_{SU(2)}^{-1}\chi_{2}~, 
\eea
whilst the vielbeins that change are\footnote{Notice that we have rotated $\hat{e}^{1}$ and $\hat{e}^{2}$ with respect to those presented in appendix 6 of \cite{Macpherson:2014eza}. We have also added some extra factors of $\lambda$ and $\lambda_{1}$ which we found to be missing.}
\bea
\hat{e}^{1}&=&-\frac{\lambda_{1}}{L Q}\left[( (L^{4}\lambda_{1}^{2}\lambda^{2}+x_{1}^{2})\cos\xi+L^{2}\lambda^{2} x_{2}\sin \xi)\ud x_{1}\right.\nonumber\\
&&\left.+x_{1}(x_{2}\cos \xi -L^{2}\lambda_{1}^{2}\sin\xi)(\ud x_{2}+L^{2}\lambda^{2}(\ud \xi+\cos\theta_{1} \ud \phi_{1}))\right]\nonumber\\%%%%%%
\hat{e}^{2}&=&-\frac{\lambda_{1}}{LQ}\left[((L^{4}\lambda^{2}\lambda_{1}^{2}+x_{1}^{2})\sin\xi-L^{2}\lambda^{2}x_{2}\cos\xi)\ud x_{1}\right.\nonumber\\ 
&&\left.+x_{1}(L^{2}\lambda_{1}^{2}\cos\xi+x_{2}\sin\xi)(\ud x_{2}+ L^{2}\lambda^{2}(\ud \xi+\cos\theta_{1}\ud \phi_{1}))\right]\nonumber\\%%%%
\hat{e}^{3}&=&-\frac{\lambda}{L Q}[x_{1}x_{2}\ud x_{1}+(L^{4}\lambda_{1}^{4}+x_{2}^{2})\ud x_{2}-L^{2} \lambda_{1}^{2}x_{1}^{2}(\ud \xi +\cos \theta_{1} \ud \phi_{1})].
\eea
One now has all the information to perform the T-duality. After both dualities the $T^{(1,1)}$ spinors become
\bea 
\chi_{1}\xrightarrow[\text{NATD}]{ }\chi_{1}\xrightarrow[~\text{T}~]{} \chi_{1}~,\quad \chi_{2}\xrightarrow[\text{NATD}]{} \Omega^{-1}_{SU(2)}\chi_{2}\xrightarrow[~\text{T}~]{} \Omega^{-1}_{U(1)}\Omega_{SU(2)}^{-1}\chi_{2}~.
\eea
One may now compute all the spinor bilinears. One finds for the scalar bilinears
\bea
A&=& 1~,\\
\sin \zeta&=&0~,\\
Z&=&0~,\\
S&=&-\frac{\lambda_{1}^{2}x_{1} \sin\theta e^{\ii \xi}}{\sqrt{W}}~.
\eea
From $S$ one finds 
\bea
\xi=-\psi~, \qquad \taul\mu=\frac{\lambda_{1}^{2} x_{1} \sin\theta }{\sqrt{W}}~.
\eea
Moreover one sees that the warp factor arises from putting the $d=10$ metric into Einstein frame and therefore we have the identification $\me^{2 \Delta}=\mu^{-1/2}=\me^{-\Phi/2}$. From this we find
\bea
\taul=L^{2}\lambda_{1}^{2}x_{1} \sin\theta_{1}~.
\eea
One is able to find the one-form bilinears $K_{5}$ and $K$ from this information by using (\ref{explicitK}) and (\ref{K5final}) and we may use this as a check for the result defined directly from the Killing spinors. Computing the one-form bilinears form the Killing spinors one finds
\bea
K&=&\frac{L \lambda\lambda_{1}^{2} e^{i \xi}}{\sqrt{W}}(i(\sin \theta \ud x_{1}+x_{1}\cos\theta \ud \theta)-x_{1}\sin\theta \ud \xi)~, \\
K_{5}&=& -\frac{L \lambda \lambda_{1}^{4} x_{1}^{2} \sin^{2}\theta}{W}\ud \xi~,\\
K_{4}&=&\frac{\lambda(-x_{1}\cos\theta_{1} \ud x_{1}-x_{2}\cos\theta_{1} \ud x_{2}+L^{4}\lambda_{1}^{4}\sin \theta_{1} \ud \theta_{1} +x_{2} \ud \phi_{1})}{L \sqrt{W}}~,\\
K_{3}&=& \frac{\lambda \lambda_{1}^{2}}{L W}[x_{1}x_{2}\sin^{2}\theta_{1} \ud x_{1}+(L^{4}\lambda_{1}^{4}+x_{2}^{2})\sin^{2}\theta_{1} \ud x_{2}+x_{1}^{2}\cos\theta_{1} \ud \phi_{1}\nonumber\\%%%%%%%
&&+\ii L^{2} \sqrt{W}(\cos\theta_{1} \ud x_{2}+x_{2}\sin \theta_{1} \ud \theta_{1} -\ud \phi_{1})].
\eea
Finally, using the redefinitions used in the classification (\ref{K5final}) and (\ref{K3final}), one recovers (\ref{Natdtau})-(\ref{Natdsigma}). The change of coordinates (\ref{changeofcoords1})-(\ref{changeofcoords4}) follows from noticing that $\phi$ can be identified with $x$ and then observing that certain combinations of $\ud x_{i}$ and $\ud \theta_{1}$ appear only. From these combinations by adding suitable functions and requiring that they are closed one recovers the change of coordinates presented.

\end{document}